\documentclass[aps,prb,twocolumn,superscriptaddress,showpacs]{revtex4-1}
\usepackage{amsmath,amssymb,graphicx}
\usepackage[hypertex,breaklinks=true,colorlinks=false]{hyperref}

\newcommand{\bol}[1]{\boldsymbol #1}
\newcommand{\PRL}[3]{Phys. Rev. Lett. {\bf #1},
\href{http://link.aps.org/abstract/PRL/v#1/e#2}{#2} (#3)}
\newcommand{\PRLp}[3]{Phys. Rev. Lett. {\bf #1},
\href{http://link.aps.org/abstract/PRL/v#1/p#2}{#2} (#3)}
\newcommand{\PRB}[3]{Phys. Rev. B {\bf #1},
\href{http://link.aps.org/abstract/PRB/v#1/e#2}{#2} (#3)}
\newcommand{\PRBp}[3]{Phys. Rev. B {\bf #1},
\href{http://link.aps.org/abstract/PRB/v#1/p#2}{#2} (#3)}
\newcommand{\PRBR}[3]{Phys. Rev. B {\bf #1},
\href{http://link.aps.org/abstract/PRB/v#1/e#2}{#2(R)} (#3)}
\newcommand{\JPSJ}[3]{J. Phys. Soc. Jpn. {\bf #1},
\href{http://jpsj.ipap.jp/link?JPSJ/#1/#2/}{#2} (#3)}

\begin{document}
\title{Mass ratio of elementary excitations in
frustrated antiferromagnetic chains with dimerization}
\author{Shintaro Takayoshi}
\author{Masaki Oshikawa}
\affiliation{Institute for Solid State Physics, University of Tokyo, 
Kashiwa, Chiba 277-8581, Japan}
\date{\today}

\begin{abstract}
Excitation spectra of $S=1/2$ and $S=1$
frustrated Heisenberg antiferromagnetic chains with
bond alternation (explicit dimerization) are studied 
using a combination of analytical and numerical methods.
The system undergoes a dimerization transition at a critical
bond alternation parameter $\delta=\delta_{\rm c}$, where
$\delta_{\rm c} = 0$ for the $S=1/2$ chain.
The SU(2)-symmetric sine-Gordon theory
is known to be an effective field theory of the system 
except at the transition point. 
The sine-Gordon theory has a SU(2)-triplet and
a SU(2)-singlet of elementary excitation, and
the mass ratio $r$ of the singlet to the triplet
is $\sqrt{3}$. 
However, our numerical calculation 
with the infinite time-evolving block decimation method shows
that $r$ depends on the frustration (next-nearest-neighbor 
coupling) and is generally different from $\sqrt{3}$.
This can be understood as an effect of marginal perturbation
to the sine-Gordon theory.
In fact, at the critical frustration separating the second-order and
first-order dimerization transitions, the marginal operator
vanishes and $r=\sqrt{3}$ holds.
We derive the mass ratio $r$ analytically 
using form-factor perturbation theory combined with
a renormalization-group analysis.
Our formula agrees well with the numerical results,
confirming the theoretical picture.
The present theory also implies that,
even in the presence of a marginally irrelevant operator,
the mass ratio approaches $\sqrt{3}$
in the very vicinity of the second-order
dimerization critical point $\delta \sim \delta_c$.
However, such a region is extremely small and would be difficult
to observe numerically.
\end{abstract}


\maketitle

\section{Introduction}

Techniques of field theory have achieved growing success 
in interpreting physical properties in low dimensional magnets. 
The achievement stems from the close connection between 
one-dimensional quantum spin models and their effective theories. 
In particular, $S=1/2$ Heisenberg antiferromagnetic (HAF) chains 
with various perturbations are important and also relevant for
experimental studies of one-dimensional magnets.
The bosonization scheme~\cite{Giamarchi} 
is useful for analyzing these systems. 
A HAF chain with bond alternation, or under a staggered field, is
described effectively by the sine-Gordon (SG) field theory. 
Elementary excitations in these systems are
a soliton, an antisoliton, and breathers 
(bound states of the soliton and antisoliton). 
Materials such as Cu benzoate~\cite{Oshikawa97,Affleck99} 
and KCuGaF$_{6}$~\cite{Umegaki11} are described by HAF 
in a staggered field, and the soliton gap calculated from SG theory 
explains well the experimental results. 
For dimerized chains, the gap formula as a function of dimerization $\delta$ 
with logarithmic correction is obtained~\cite{Affleck89}: 
$\delta^{2/3}/|\log\delta|^{1/2}$, 
or it can also be represented as an effective power-law form 
with a renormalized exponent which deviates from 2/3.~\cite{Singh99} 
Refined logarithmic correction is given in Ref.~\onlinecite{Orignac04}.
Dimerized spin chains are an appropriate model for spin-Peierls materials 
such as CuGeO$_{3}$~\cite{Hase93} or Ni compounds.~\cite{Hagiwara98}

There are also a number of numerical studies on
the frustrated HAF chain with next-nearest-neighbor coupling.
We consider the Hamiltonian
\begin{equation}
 {\cal H}= J \sum_{j}\left[\{1+(-1)^{j}\delta\}
                   \bol{S}_{j}\cdot\bol{S}_{j+1}
                 +\alpha \bol{S}_{j}\cdot\bol{S}_{j+2}\right],
\label{eq.model}
\end{equation}
where $J>0$.
The next-nearest-neighbor coupling $\alpha \geq 0$ introduces
frustration.

This model exhibits a dimerization transition at $\delta=\delta_{\rm c}$.
For $S=1/2$, the transition point is always $\delta_{\rm c}=0$,
since the Lieb-Schultz-Mattis theorem implies either gapless
excitations or two-fold degeneracy of the ground states at $\delta=0$.
In fact, on the undimerized line $\delta=0$,
there exists a critical frustration parameter
$\alpha_{\rm c} \sim 0.2411$.~\cite{Haldane-NNN,Okamoto92} 
For $\alpha < \alpha_{\rm c}$ the system is a gapless Tomonaga-Luttinger
Liquid (TLL); that is, the dimerization transition at
$\delta = \delta_{\rm c} =0$ is of second order.
In contrast, for $\alpha > \alpha_{\rm c}$, the ground state
is doubly degenerate, exhibiting a spontaneous dimerization.
This implies a first-order dimerization transition
at $\delta=\delta_{\rm c} = 0$.

For $S=1$, on the other hand, $\delta=0$ (for a small $\alpha$)
belongs to the Haldane phase and does not represent a transition line.
Instead, a dimerization transition between the Haldane phase
and the dimerized phase
occurs~\cite{AffleckHaldane,Kato94,Pati96,Kolezhuk96}
at a finite $\delta_{\rm c}$,
which depends on the frustration $\alpha$.
Although the shape of the phase diagram is thus different,
the topology of the phase diagram is rather similar to that
for $S=1/2$.
In fact, also for $S=1$, there is a critical frustration
$\alpha_{\rm c}$; the transition is second order with the critical
point described by a TLL for $\alpha<\alpha_{\rm c}$, and
first order for $\alpha>\alpha_{\rm c}$.

In the neighborhood of the gapless TLL line,
the system acquires a small excitation gap,
and would be described by the SG theory.
Since our model~\eqref{eq.model} is SU(2)-invariant,
the SG theory should also have SU(2)-symmetry.
As a consequence,
the mass ratio $r$ of the second lowest breather 
to the soliton should be $\sqrt{3}$.

However, numerical results for $S=1/2$ chains~\cite{Bouzerar98} 
show that $r$ generally does not agree with the SG theory
prediction $\sqrt{3}$.
While $r$ depends only weakly on $\delta$,
it does vary as a function of $\alpha$.
Only near the critical frustration $\alpha=\alpha_{\rm c}$ 
does $r$ agree with the SG prediction $\sqrt{3}$.
In Ref.~\onlinecite{Bouzerar98}, it was pointed out that
a marginal operator exists as a perturbation to the
SG theory, and it would shift $r$ from $\sqrt{3}$.
However, how exactly the mass ratio $r$ is affected
by the marginal operator was not clarified.

The effect of the marginal perturbation to the SG theory
on the mass ratio was discussed in terms of
form-factor perturbation theory (FFPT) 
in Ref.~\onlinecite{Controzzi04}.
However, the theoretical prediction has not been tested.
The mass ratio in the $S=1$ case has also
never been studied numerically.

In this paper,
we study numerically the mass ratio of elementary excitations
and the ground phase diagram of the frustrated HAF
with bond alternation~\eqref{eq.model} for
both $S=1/2$ and $S=1$.
We employ the recently developed infinite time-evolving
block decimation (iTEBD) method,~\cite{Vidal07} 
which allows high-precision calculation of infinitely long
chains.
The masses of elementary excitations are obtained from
the asymptotic behavior of equal-time correlation
functions, instead of extrapolation of the finite-size
energy spectrum.
We confirm previous results when they are available,
and we obtain the mass ratio for $S=1$ as a new result.
Furthermore, we derive an explicit formula for
the mass ratio $r$ as a function of $\delta$ and $\alpha$,
by combining FFPT and renormalization-group analysis.
This agrees well with the numerical results for both $S=1/2$
and $S=1$.
Thus both cases are understood in terms of
the unified framework of the SG theory with
a marginal perturbation.

This paper is organized as follows.
In Secs.~\ref{sec.bosonization} and \ref{sec.NLSM} respectively,
we review direct bosonization of the $S=1/2$ chain
and derivation of the SG theory for general $S$ case
via the $O(3)$ nonlinear sigma model (NLSM).
In Secs.~\ref{sec.spin1/2} and \ref{sec.spin1}, we
present numerical study on the mass ratio
and phase diagram, respectively for $S=1/2$ and $S=1$.
We then discuss the mass ratio analytically based on FFPT
and compare the theoretical formula with the numerical results
in Sec.~\ref{sec.FFPT}.
Sec.~\ref{sec.conclusion} is devoted to conclusions.

\section{Bosonization}
\label{sec.bosonization}

We first review the bosonization of a spin-$1/2$ chain. 
Spin operators are represented as 
\begin{align}
 S_{j}^{z}&=\frac{a}{\pi}\partial_{x}\phi
           +a_{1}(-1)^{j}\cos(2\phi)+\cdots\nonumber\\
 S_{j}^{+}&={\rm e}^{{\rm i}\theta}\left[
            b_{0}(-1)^{j}+b_{1}\cos(2\phi)+\cdots\right],\nonumber
\end{align}
where dual boson fields $\phi,\theta$ satisfy the commutation relation 
$[\phi(x),\theta(x')]=-{\rm i}\pi\vartheta(x-x')$ 
($\vartheta(x-x')$ is the step function) 
with $x=ja$ ($a$ is lattice spacing). 
$\phi$ and $\theta$ have periodicity 
$\phi\sim\phi+\pi,\theta\sim\theta+2\pi$. 
Effective Hamiltonian of XXZ chain with dimerization is written 
with $\phi$ and $\theta$ as~\cite{Giamarchi} 
\begin{align}
 {\cal H}_{\rm eff}&=\frac{u}{2\pi}\int{\rm d}x
                          [K^{-1}(\partial_{x}\phi)^{2}+K(\partial_{x}\theta)^{2}]
                          \nonumber\\
                   &+\frac{2g_{1}}{(2\pi a)^{2}}\int{\rm d}x\cos(2\phi)
                    +\frac{2g_{2}}{(2\pi a)^{2}}\int{\rm d}x\cos(4\phi).
                    \label{eq:Hami_eff}
\end{align}
Irrelevant terms are omitted here. 
$u$ and $K$ denote spinon velocity and the Luttinger parameter, respectively. 
At the SU(2)-symmetric Heisenberg point, $u=\pi a/2$ and $K=1/2$. 
Since the operator ${\rm e}^{{\rm i}q\phi(x)}$ has scaling dimension 
$Kq^{2}/4$, the $\cos(2\phi)$ term is relevant 
while the $\cos(4\phi)$ term becomes marginal. 
The $g_{1}$ term arises from the bond alternation (i.e. dimerization).
$g_{2}$ is known to decrease with increasing $\alpha$ and vanish at 
$\alpha = \alpha_{\rm c}$ where the transition from TLL
to the self-dimerized phase happens. 
Thus, coupling constants $g_{1}$ and $g_{2}$ are proportional to 
$\delta$ and $\alpha - \alpha_{\rm c}$, respectively. 
When $g_{1}\neq 0$ and $g_{2}=0$, 
\eqref{eq:Hami_eff} is equivalent to the SG model. 
It is an exactly solved model, and 
the excitation spectrum is obtained.~\cite{Essler98,Dashen75} 
There appear three types of elementary particles: 
a soliton, a corresponding antisoliton, and breathers. 
The number of breathers is $[2/K-1]$, where $[A]$ stands for the integer 
part of $A$. The mass of soliton $M_{\rm S}$ and 
the $n$-th lightest breather $M_{{\rm B}_n}$ are related through the formula
\begin{equation}
 M_{{\rm B}_n}=2 M_{\rm S} \sin\left(\frac{n\pi}{4/K-2}\right), 
              \quad n=1,\cdots,[2/K-1]. \label{eq:mass_relation}
\end{equation}
According to \eqref{eq:mass_relation}, 
in HAF chain with dimerization ($K=1/2$), 
the soliton, the antisoliton and the first breather form triplet 
while the second breather is a singlet which has $\sqrt{3}$-times 
as large mass as the triplet. 
Although the degeneracy of the triplet is protected thanks to SU(2)-symmetry, 
the mass ratio of singlet to triplet $r\equiv M_{{\rm B}_2}/M_{\rm S}$ 
is subject to correction caused by the marginal term $g_2$.

\section{SG theory via nonlinear sigma model}
\label{sec.NLSM}

$S>1/2$ chains may be bosonized by introducing
Hund coupling to $2S$ chains of spin-$1/2$.
Each chain is bosonized separately, resulting in
a theory of interacting $2S$ boson fields.~\cite{Schulz86} 
In the low-energy limit, however, one of the linear
combinations of the boson fields becomes important.
The SG theory (or TLL) would emerge as an effective theory
of this linear combination.

However, it is rather cumbersome to pursue this explicitly.
As an alternative, the SG theory can also be derived
from the $O(3)$ non-linear sigma model (NLSM).
The $O(3)$ NLSM was derived in the semi-classical,
large-$S$ limit of the HAF chain.
Nevertheless, it proved to be a useful effective theory
even for $S=1$.

Let us define fields $\bol{n}(x)$ and $\bol{l}(x)$ 
by $\bol{S}_{j}/S\sim (-1)^{j}\bol{n}(x)+\bol{l}(x)$. 
Then the spin-$S$ HAF chain with bond alternation~\eqref{eq.model}
can be generally mapped to the $O(3)$ NLSM
\begin{equation}
{\cal A}_\theta=\frac{1}{2g}\int{\rm d}\tau{\rm d}x
                \left\{v(\partial_{x}\bol{n})^{2}
                      +\frac{1}{v}(\partial_{\tau}\bol{n})^{2}\right\}
               +{\rm i}\theta T, \nonumber
\end{equation}
where $g=2/S$ is some coupling constant and $v=2JS$ is the spin-wave velocity.
$T=\frac{1}{4\pi}\int{\rm d}\tau{\rm d}x\bol{n}\cdot
\partial_{x}\bol{n}\times\partial_{\tau}\bol{n}$ represents 
the integer-valued topological charge and $\theta=2\pi S (1+\delta)$. 
For the moment, let us assume that there is no frustration, $\alpha=0$.

$O(3)$ NLSM is known to be integrable~\cite{Zam2_1979,Zam2_1992}
at $\theta=0$ and $\pi$. At $\theta \equiv 0 \mod{2\pi}$,
the excitation consists of a triplet of massive particles. 
In contrast, the theory is massless at $\theta \equiv \pi \mod{2\pi}$
and the infrared fixed point is a SU(2)$_{1}$ Wess-Zumino-Witten model, 
a conformal field theory (CFT) with central charge $c=1$. 
This is merely the TLL at the SU(2)-symmetric point $K=1/2$.

When bond alternation is absent ($\delta=0$),
the system is massless ($\theta=\pi$) if $S$ is a half-odd-integer,
while it is massive ($\theta=0$) if $S$ is an integer.
This is the celebrated Haldane conjecture,~\cite{Haldane83} 
which is now established by intensive analytical, numerical, and
experimental studies.

It is also interesting to consider the effect of
bond alternation $\delta$.
By changing $\delta$ from $-1$ to $1$, namely from
the completely dimerized limit to the opposite completely dimerized limit,
$\theta$ passes the critical point, $\pi \mod{2\pi}$, $2S$ times.
Thus, on $-1 < \delta < 1$,
there are $2S$ successive phase transitions.~\cite{AffleckHaldane}
This could be understood as successive spontaneous breaking
and restoration of hidden symmetry,~\cite{Oshikawa92} or more generally,
symmetry-protected topological phase transitions.~\cite{Pollmann2010,Pollmann2012}

For $S=1/2$, the transition occurs only at $\delta=0$,
consistently with the direct bosonization analysis.
For $S=1$, there are two transitions which separate the
Haldane phase around $\delta=0$ from the dimerized phases.
The critical points are, according to the above argument,
given by $\delta = \pm \delta_{\rm c} = \pm 1/2$.
However, in reality, the location of the critical points
is renormalized.
It was shown~\cite{Kato94} numerically that
$\delta_{\rm c}\sim 0.25J $. 

As discussed above, the critical point is described by
the SU(2)-symmetric TLL with $K=1/2$.
By considering the possible perturbations to the TLL,
the effective theory near the critical
point $\delta=\delta_{\rm c}$ is determined~\cite{Controzzi04}
to be the SG theory with marginal perturbation~\eqref{eq:Hami_eff}, 
which was derived previously for $S=1/2$ by direct bosonization.
Thus, the same theory~\eqref{eq:Hami_eff}
should describe the neighborhood of dimerization transitions
for any $S$.
In the following, we shall investigate the systems
with $S=1/2$ and $S=1$ numerically, and verify
this universality.

\section{Mass ratio and phase diagram for $S=1/2$}
\label{sec.spin1/2}

We study the excitation spectrum of the system numerically,
and we focus in particular on the change of $r$
due to the marginal term.
We adopt a new strategy to extract the excitation spectrum
from the equal-time correlation function obtained by iTEBD, shown as follows.

A single-particle excitation in the SG model can be parameterized by
the rapidity $\theta$, which defines its energy and wave number as 
$M_{0}\cosh\theta$ and $(M_{0}/u)\sinh\theta$, respectively 
($M_{0}$ is the mass of the particle). 
The one-particle form factor of operator ${\cal O}$ 
is specified by $\theta$ and the kind of particle $a$ as 
$F_{\cal O}(\theta, a)\equiv\langle 0|{\cal O}|\theta, a\rangle$. 
${\cal O}$ represents an operator which creates 
the single soliton, the antisoliton, or the breather. 
We can calculate the equal-time correlation function
$ \langle{\cal O}(r){\cal O}(0)\rangle - \langle{\cal O}(r)\rangle\langle{\cal O}(0)\rangle$
by inserting the resolution of the identity $\hat{1}=\sum_{n=0}^\infty P_{n}$ 
where $P_{n}$ is the projection operator defined as 
$P_{0}=|0\rangle \langle 0 |$ and 
$P_{n}=\frac{1}{n!}\sum_{a_1,\cdots,a_n}\int\frac{\prod_{j}{\rm d}\theta_j}{(4\pi)^{n}}
       |\theta_{1}, a_{1}; \cdots; \theta_{n}, a_{n}\rangle
       \langle \theta_{1}, a_{1}; \cdots; \theta_{n}, a_{n}|$ ($n \geq 1$).
Then, the leading order of the correlation function~\cite{Furuya11} is
\begin{align}
 \langle{\cal O}(r){\cal O}(0)\rangle&-\langle{\cal O}(r)\rangle\langle{\cal O}(0)\rangle
 \nonumber\\ 
 &\approx\int\frac{{\rm d}\theta}{4\pi}
             {\rm e}^{{\rm i}M_{0}r\sinh\theta/u}|F_{\cal O}(\theta, a)|^{2}.\nonumber
\end{align}
In the limit of $l\to\infty$, it is calculated to be~\cite{Kuzmenko09} 
\begin{equation}
 \langle{\cal O}(l){\cal O}(0)\rangle-\langle{\cal O}(l)\rangle\langle{\cal O}(0)\rangle
 =(A(-1)^{l}+B)\frac{{\rm e}^{-l/\xi}}{\sqrt{l}}\label{eq:correl_fit}
\end{equation}
consisting of a staggered and uniform part. 
We suppose that the effect of the marginal $\cos(4\phi)$ term is renormalized 
into mass $M_{0}$ and constants $A,B$. 

\begin{figure}[b]
\begin{center}
\includegraphics[width=0.3\textwidth]{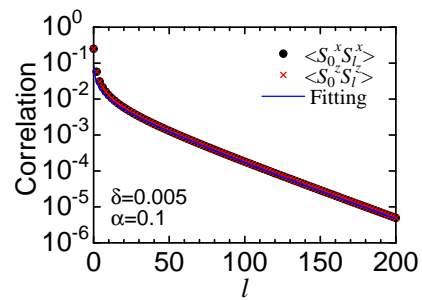}
\end{center}
\caption{(Color online) 
Correlation functions $\langle S_{0}^{x}S_{l}^{x}\rangle$ and 
$\langle S_{0}^{z}S_{l}^{z}\rangle$ calculated with the iTEBD method. 
The solid line represents the fitting with the function $C{\rm e}^{-l/\xi}/\sqrt{l}$.}
\label{fig:Correlation}
\end{figure}

In this way, the mass can be extracted from the correlation function,
which we calculate with the iTEBD method.
The truncation dimension, the number of conserved states in 
evolution, is fixed to be 200, 
large enough for the iTEBD calculation in gapped systems.
$\langle S_{0}^{x}S_{l}^{x}\rangle$, 
$\langle S_{0}^{y}S_{l}^{y}\rangle$, 
$\langle S_{0}^{z}S_{l}^{z}\rangle$ and 
$\langle (\bol{S}_{0}\cdot \bol{S}_{1})(\bol{S}_{l}\cdot \bol{S}_{l+1})\rangle$ 
are fitted with $C{\rm e}^{-l/\xi}/\sqrt{l}$ 
for sufficiently large and even $l$. 
$C(=A+B)$ and $\xi$ are fitting parameters. 
Then we can obtain the mass of the soliton, the antisoliton, and the first and 
second breathers, respectively, through the relation $M=u/\xi$. 
Note that $M$ is a renormalized mass. 
While the value $u$ for $\alpha=0$ is obtained exactly from the Bethe ansatz, 
it cannot be for $\alpha \neq 0$.
Yet, the value of $u$ is not needed to calculate a mass ratio. 

Since $S_{\rm tot}^{z}\equiv\sum_{j}S_{j}^{z}$ commutes with the Hamiltonian, 
$S_{\rm tot}^{z}$ is a good quantum number. 
The ground state is in $S_{\rm tot}^{z}=0$ Hilbert space, 
and the soliton (antisoliton) is an excitation to 
the lowest energy level in $S_{\rm tot}^{z}=1(-1)$ Hilbert space. 
Hence, their mass corresponds to the inverse correlation length 
of the operator changing $S_{\rm tot}^{z}$ 
by $\pm 1$~\cite{Takayoshi10}, 
i.e. $\langle S_{0}^{x}S_{l}^{x}\rangle=\langle S_{0}^{y}S_{l}^{y}\rangle$. 
On the other hand, the first breather is the lowest excitation in 
$S_{\rm tot}^{z}=0$ Hilbert space; it corresponds to 
$\langle S_{0}^{z}S_{l}^{z}\rangle$. 
In the case of the antiferromagnetic XXZ model, 
the mass of the soliton / antisoliton and the first breather is different. 
For a Heisenberg chain, however, SU(2)-symmetry requires 
$\langle S_{0}^{x}S_{l}^{x}\rangle=\langle S_{0}^{y}S_{l}^{y}\rangle 
=\langle S_{0}^{z}S_{l}^{z}\rangle$, 
which indicates that the mass of the soliton, the antisoliton, 
and the first breather is all the same, 
and these three particles constitute a triplet. 
The second breather has to be a singlet, and the operator corresponding 
to it does not change $S_{\rm tot}^{z}$. 
The most relevant operator with such properties is $\bol{S}_{j}\cdot\bol{S}_{j+1}$, 
and we expect that the second breather corresponds to 
$\langle (\bol{S}_{0}\cdot \bol{S}_{1})(\bol{S}_{l}\cdot \bol{S}_{l+1})\rangle$. 

\begin{figure}
\begin{center}
\includegraphics[width=0.3\textwidth]{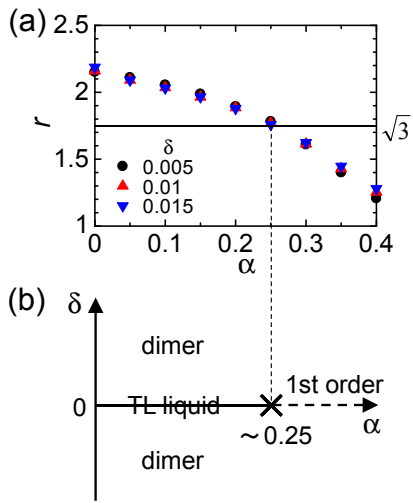}
\end{center}
\caption{(Color online) 
(a) Triplet-singlet mass ratio $r$ as a function 
of $\alpha$ and $\delta$ in the $S=1/2$ bond-alternating
chain with frustration. 
(b) Phase diagram of the $S=1/2$ bond-alternating
chain with frustration.
Solid and dashed lines represent second order 
(TLL, $c=1$ CFT) and first order transition, respectively.
Universality class of transition changes at 
$\alpha \sim 0.25$, where $r$ becomes $\sqrt{3}$.}
\label{fig:S1ov2RatioPD}
\end{figure}

An example of fitting for correlation functions is shown in Fig.~\ref{fig:Correlation}.
$\langle S_{0}^{x}S_{l}^{x}\rangle$ and $\langle S_{0}^{z}S_{l}^{z}\rangle$ 
calculated with the iTEBD method are equal up to eight digits, 
which is consistent with the SU(2)-symmetry. 
The solid line represents the fitting with the function $C{\rm e}^{-l/\xi}/\sqrt{l}$. 
The correlation functions are well fitted with the function. 

We show numerically the calculated mass ratio $r$ as a function 
of $\alpha$ and $\delta$ in Fig.~\ref{fig:S1ov2RatioPD}(a). 
$r$ is larger than 2 for
$\alpha=0$ (nonfrustrated HAF chain with bond alternation)
and decreases with increasing $\alpha$. 
$r$ becomes $\sqrt{3}$ at $\alpha \sim 0.25$. 
It is very close to $\alpha = 0.2411$, 
where the transition from TLL to the self-dimerized phase 
happens without bond alternation, and 
the marginal $\cos(4\phi)$ term vanishes.~\cite{Okamoto92} 
This result indicates that the deviation of $r$ from $\sqrt{3}$ 
is attributed to the effect of the marginal term. 
While $r$ is subject to correction as $\alpha$ moves away from this point, 
its $\delta$ dependence is quite small. 

A similar result was obtained through a gap evaluation by 
exact diagonalization.~\cite{Bouzerar98} 
However, the mechanism of the variation of $r$
has not been made clear. 
We will theoretically analyze the dependence of $r$ on the
frustration $\alpha$ later in Sec.~\ref{sec.FFPT}.
The $\alpha$-$\delta$ phase diagram is shown in Fig.~\ref{fig:S1ov2RatioPD}(b).
Note that the universality class of the transition from positive to negative 
$\delta$ is of $c=1$ CFT for $\alpha<0.25$ 
and of first order for $\alpha>0.25$.~\cite{Chitra95} 

\begin{figure}
\begin{center}
\includegraphics[width=0.3\textwidth]{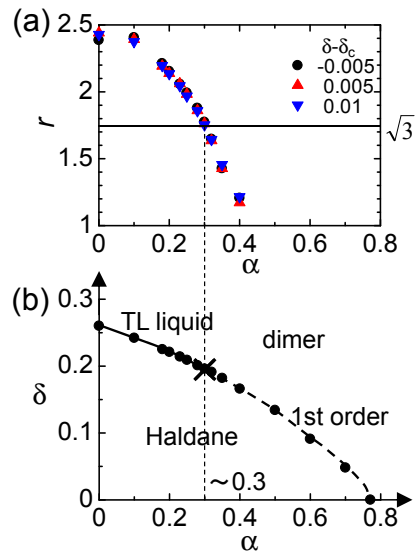}
\end{center}
\caption{(Color online) 
(a) Triplet-singlet mass ratio $r$ as a function 
of $\alpha$ and $\delta$ in the $S=1$ bond-alternating
chain with frustration. 
The transition point $\delta_{\rm c}$ corresponds to 
circles in (b). 
(b) Phase diagram of the $S=1$ bond-alternating
chain with frustration.
The solid and dashed lines, which represent the second order 
(TLL, $c=1$ CFT) and first order transition, respectively, 
are guides for the eye. 
The circles show transition points determined from 
Fig.~\ref{fig:S1CorDimer}. 
The universality class of the transition changes approximately at 
$\alpha=0.3$, where $r$ becomes $\sqrt{3}$.}
\label{fig:S1RatioPD}
\end{figure}

\section{Mass ratio and phase diagram for $S=1$}
\label{sec.spin1}

\begin{figure}
\begin{center}
\includegraphics[width=0.3\textwidth]{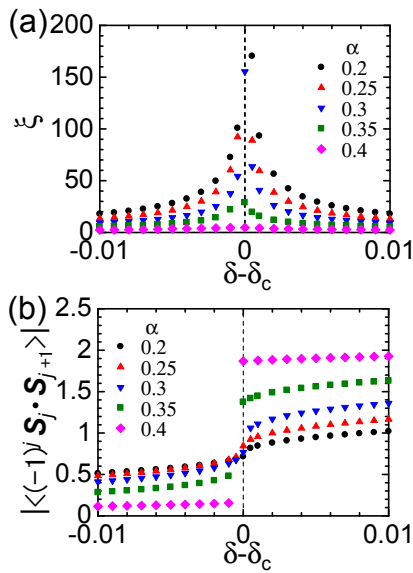}
\end{center}
\caption{(Color online) 
(a) Correlation length $\xi$ as a function of 
$\delta-\delta_{\rm c}$ and $\alpha$. 
$\xi$ diverges when the transition is second order.
(b) Dimerization order parameter 
$|\langle(-1)^{j}S_{j}\cdot S_{j+1}\rangle|$ as a function of 
$\delta-\delta_{\rm c}$ and $\alpha$. 
$|\langle(-1)^{j}\bol{S}_{j}\cdot\bol{S}_{j+1}\rangle|$ jumps 
when the transition is first order.
Therefore, $\delta_{\rm c}$ can be determined from the divergence of $\xi$ or 
the jump of $|\langle(-1)^{j}\bol{S}_{j}\cdot\bol{S}_{j+1}\rangle|$.}
\label{fig:S1CorDimer}
\end{figure}

Next, we numerically investigate the excitation spectrum
and the phase diagram of the $S=1$ HAF chain with dimerization and frustration. 
The method for evaluating particle mass is the same as 
for the $S=1/2$ chain. 
As can be seen in Fig.~\ref{fig:S1RatioPD}, 
when $\alpha$ is small enough, 
$r$ is always larger than 2 at least in $|\delta-\delta_{\rm c}|\geq 0.005$ 
($\delta_{\rm c}$ can be determined from the divergence of $\xi$ or 
the jump of $|\langle(-1)^{j}\bol{S}_{j}\cdot\bol{S}_{j+1}\rangle|$ 
as explained in the following part. See Fig.~\ref{fig:S1CorDimer}), 
and does not depend much on $\delta$. 
Since particles heavier than $2M_{\rm S}$ become resonance, 
the second breather cannot be a stable particle 
even in the vicinity of $\delta_{\rm c}$. 
The above result again seems inconsistent with 
the prediction in Ref.~\onlinecite{Controzzi04}. 

The deviation of $r$ from $\sqrt{3}$ would be attributed to the 
existence of the marginal term as in the spin-$1/2$ chain. 
We introduce the next-nearest-neighbor coupling $\alpha$ 
in order to confirm it. 
As shown in Fig.~\ref{fig:S1RatioPD}(a), 
$r$ decreases with increasing $\alpha$ and becomes $\sqrt{3}$ 
around $\alpha=0.3$. 
The transition point $\delta_{\rm c}$ from the Haldane phase 
to the dimerized phase also decreases, 
which is natural because next-nearest-neighbor coupling 
favors the dimerized phase. 
Fig.~\ref{fig:S1CorDimer} shows the behavior of 
correlation length $\xi$ and dimerization order parameter 
$|\langle(-1)^{j}\bol{S}_{j}\cdot\bol{S}_{j+1}\rangle|$ 
near $\delta_{\rm c}$. 
$\xi$ diverges at $\delta_{\rm c}$ for $\alpha \lesssim0.3$, 
which is not the case for $\alpha >0.3$. 
In addition, $|\langle(-1)^{j}\bol{S}_{j}\cdot\bol{S}_{j+1}\rangle|$ 
jumps at $\delta_{\rm c}$ for $\alpha >0.3$ 
while the variation is continuous for $\alpha \lesssim0.3$. 
These results indicate that the universality class of transition 
at $\delta_{\rm c}$ changes from $c=1$ CFT to first order 
when $\alpha$ goes beyond 0.3. 
From the viewpoint of field theory, the $\cos(4\phi)$ term changes from 
a marginally irrelevant to a marginally relevant operator 
at this point. 
The situation is very analogous to the spin-$1/2$ case. 
The $\alpha$-$\delta$ phase diagram is summarized 
in Fig.~\ref{fig:S1CorDimer}(b). 
It is consistent with Ref.~\onlinecite{Pati96}. 
The transitions along the lines $\delta=0$ and $\alpha=0$ 
are studied in Refs.~\onlinecite{Kolezhuk96} 
and \onlinecite{Kato94}, respectively. 

\section{Mass ratio from the form-factor perturbation theory}
\label{sec.FFPT}

Now let us discuss the variation of the mass ratio $r$
theoretically. 
In Ref.~\onlinecite{Controzzi04},
the $\delta$ dependence of $r$ was discussed as follows.
The excitation structure at the very vicinity of
$\delta=\delta_{\rm c}$ would be described by
the pure SG theory without the marginal perturbation;
$r$ is then equal to $\sqrt{3}$. 
On the other hand, $O(3)$ NLSM with $\theta=0$ also has triplet 
lowest excitation, which is smoothly connected to the triplet 
in the SG model thanks to SU(2)-symmetry, 
but does not have the second breather. 
Therefore, $r$ increases as $\delta$ decreases 
from $\delta_{\rm c}$ to 0, and it exceeds 2 at some point. 
This argument was further augmented by a
FFPT calculation in terms of the marginal perturbation.

However, their predictions\cite{Controzzi04} do not
seem to be consistent with numerical results.
In the absence of frustration $\alpha$,
$r$ is substantially larger than $\sqrt{3}$ even
when $\delta$ is closest to $\delta_{\rm c}$ within
the precision of the numerical calculations.
This already contradicts the picture presented in 
Ref.~\onlinecite{Controzzi04}.
Moreover, the effect of the frustration $\alpha$ was
not discussed.

Here, we will improve the FFPT by supplementing it with
a RG analysis.
Let us define a dimensionless coupling constant 
$y_{2}\equiv g_{2}/(\pi u)$. 
With the FFPT of the marginal operator in the SG theory,
mass corrections arising from the marginal term $y_2$ 
to the triplet and the singlet,
which we denote, respectively, as $\Delta M_t$ and $\Delta M_s$,
were found\cite{Controzzi04} to be
\begin{equation}
\begin{aligned}
\Delta M_t & =  4 \sqrt{3} y_2, \\
\Delta M_s & = 12 \sqrt{3} y_2.
\end{aligned}
\label{eq.correction}
\end{equation}
Here, we argue that the renormalized coupling constant
should be used for $y_2$.
In the following, we derive the renormalized form of $y_{2}$. 
Since the system has SU(2)-symmetry, $y_{2}$ is renormalized 
according to the Kosterlitz-Thouless renormalization equation~\cite{Kosterlitz74,Affleck89}, 
\begin{equation}
 \frac{{\rm d}y_{2}}{{\rm d}s}=y_{2}^{2}.\label{eq:KT_renorm}
\end{equation}
The solution of \eqref{eq:KT_renorm} is $y_{2}=-1/(s+{\rm Const.})$. 
$y_{2}$ becomes a function of energy scale 
by the parametrization $s=\ln(E/\Lambda)$ 
($\Lambda$ is the infrared cutoff) as follows
\begin{equation}
 y_{2}(E)=\frac{1}{\ln(\Lambda'/E)}.\nonumber
\end{equation}
Constant $\Lambda'$ can be fixed from the condition that 
bare $y_{2}$ corresponds to the original spin chain, 
where the energy scale is of order of $J$, i.e., 
$y_{2}(E\sim J)=C_{1} (\alpha_{\rm c} - \alpha )$, 
where $C_{1}$ is a non-universal positive constant. 
Therefore, the renormalized form of $y_{2}$ is
\begin{equation}
 y_{2}(E)=\frac{1}{\ln(J/E)+
\frac{1}{C_{1} (\alpha_{\rm c}-\alpha)}} .
\nonumber
\end{equation}
When the system is renormalized until the energy scale is equal to 
the soliton mass, $y_{2}$ becomes $y_{2}(M_{\rm S})$. 
From eq.~\eqref{eq.correction}, the mass ratio $r$ is 
\begin{equation}
 r=\frac{\sqrt3+\frac{12\sqrt{3}}{
\ln(J/M_{\rm S})+1/(C_{1}(\alpha_{\rm c}-\alpha))}}
            {1+\frac{ 4\sqrt{3}}{
\ln(J/M_{\rm S})+1/(C_{1}(\alpha_{\rm c}-\alpha))}}.
 \label{eq:Ratio_renorm}
\end{equation}

\begin{figure}
\begin{center}
\includegraphics[width=0.48\textwidth]{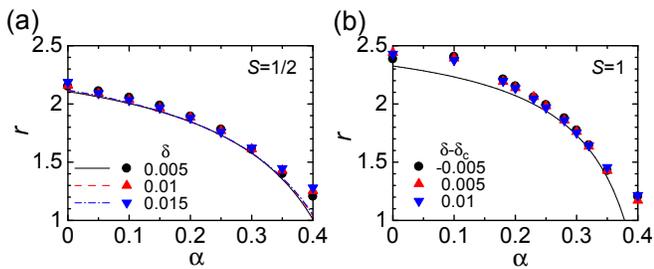}
\end{center}
\caption{(Color online) Triplet-singlet mass ratio $r$ as 
a function of $\delta$ and $\alpha$. 
(a) The case of $S=1/2$. 
The circle, triangle, and down-pointing triangle represent 
numerically obtained $r$ for $\delta=0.005$, 0.01 and 0.015, 
respectively. The solid, dashed and dashed-dotted lines are 
Eq.~\eqref{eq:Ratio_renorm} with $C_{1}=0.3$.
$M_{\rm S}$ is a function of $\delta$ and $\alpha$, 
and $M_{\rm S}$ for $\alpha=0$ is used here.
(b) The case of $S=1$. 
The circle, triangle and down-pointing triangle represent 
numerically obtained $r$ for $\delta-\delta_{\rm c}=-0.005$, 
0.005 and 0.01, respectively. 
$\delta_{\rm c}$ is determined from Fig.~\ref{fig:S1CorDimer}.
The solid line is Eq.~\eqref{eq:Ratio_renorm} 
with $M_{\rm S}=0.1J$ and $C_{1}=0.6$.}
\label{fig:RatioFitting}
\end{figure}

A fitting of the numerical results with the function~\eqref{eq:Ratio_renorm} 
is shown in Fig.~\ref{fig:RatioFitting}.
The only fitting parameter is the non-universal constant $C_{1}$. 
For $S=1/2$ chain, 
we use an excitation gap with $\alpha=0$ as the value of $M_{\rm S}$ 
since the value of $M_{\rm S}$ can be estimated through 
$M=u/\xi$, where $u=\pi J a/2$. 
The solid, dashed and dashed-dotted lines in 
Fig~\ref{fig:RatioFitting} (a) are Eq.~\eqref{eq:Ratio_renorm} 
with $C_{1}=0.3$ for $\delta=0.005$, 0.01 and 0.015, respectively. 
The variation of Eq.~\eqref{eq:Ratio_renorm} by changing $\delta$ 
is quite small since the only $\delta$-dependent variable is $M_{S}$ 
and it is present only inside a logarithm.
It is difficult to estimate $M_{\rm S}$ with
good precision for $S=1$ 
because the value of $u$ is not known. 
However, as we have discussed, the $M_{\rm S}$ dependence
is rather weak in Eq.~\eqref{eq:Ratio_renorm}.
Thus, in a practical range to compare with the numerical results,
we can set $M_{\rm S}/J=0.1$. 
Equation~\eqref{eq:Ratio_renorm} with $C_{1}=0.6$ is shown as a solid line 
in Fig~\ref{fig:RatioFitting} (b).
The fitting curves agree well with numerical data for both $S=1/2$
and $S=1$, in the vicinity of $\alpha = \alpha_{\rm c}$, where the
marginal perturbation is small.
The deviation away from the theory~\eqref{eq:Ratio_renorm}
can be attributed to higher-order correction 
in both FFPT and the renormalization equation. 

Let us come back to the argument in Ref.~\onlinecite{Controzzi04}.
As we have seen, their idea that $r$ evolves from $\sqrt{3}$
as $\theta$ is changed from $\pi \mod{2\pi}$,
does not seem to agree with the numerical results.
On the other hand, however,
where the dimerization transition is second order ($\alpha<\alpha_{\rm c}$),
the marginal operator is marginally irrelevant.
Thus, in the limit $\theta \rightarrow \pi \mod{2\pi}$
($\delta \rightarrow \delta_{\rm c}$ in our spin-chain model),
the SG theory without the marginal operator becomes exact,
and $r=\sqrt{3}$ should follow.
In this sense, their idea is still qualitatively correct.
However, the marginally irrelevant operator is renormalized to zero
very slowly (logarithmically), and thus the mass scale must
be exponentially small in order to probe this regime.
This can indeed be seen in the logarithmic dependence of $r$
on the soliton mass $M_{\rm S}$ in eq.~\eqref{eq:Ratio_renorm}.
Thus, for $\alpha<\alpha_{\rm c}$, the mass ratio deviates
very quickly from $r=\sqrt{3}$, as $\delta$ is
shifted from the critical point $\delta_{\rm c}$.
As a consequence, it would be impractical to
observe this behavior numerically.

\section{Conclusion}
\label{sec.conclusion}

We have investigated the excitation spectrum of $S=1/2$ and $1$ 
frustrated HAF chains with dimerization $\delta$. 
To evaluate particle mass $M=u/\xi$, 
we calculate the corresponding correlation function numerically 
and extract the correlation length by using 
a fitting function $C{\rm e}^{-l\xi}/\sqrt{l}$ 
for a range of large enough and even $l$. 
The ratio $r$ of the singlet (the second breather) to 
the triplet (soliton, antisoliton and the first breather) is expected to be 
$\sqrt{3}$ from bosonized SG effective field theory, 
but $r$ is subject to correction from a marginal term. 
$r=\sqrt{3}$ is recovered at the critical next-nearest-neighbor 
coupling $\alpha=\alpha_{\rm c}$,
for which the marginal term vanishes.
At $\alpha = \alpha_{\rm c}$,
the dimerization transition
with varying $\delta$ changes from
second order, with the critical behavior described by
$c=1$ CFT, to first order. 
We give $\delta$ and $\alpha$ dependences of $r$ in 
Eq.~\eqref{eq:Ratio_renorm} through FFPT and RG analysis.
$r$ obtained by the iTEBD method is well fitted by 
Eq.~\eqref{eq:Ratio_renorm}. 
Our analysis indicates that, for $\alpha<\alpha_{\rm c}$,
the mass ratio $r$ asymptotically approaches 
$\sqrt{3}$ when $\delta \rightarrow \delta_{\rm c}$, consistently
with the argument in Ref.~\onlinecite{Controzzi04}.
However, this asymptotic behavior occurs only
for exponentially small $|\delta - \delta_{\rm c}|$,
and could not be observed in numerical studies
in the literature and in the present work.

Finally, we comment on the general-$S$ case. 
When the dimerization $\delta$ is changed from 1 to $-1$, 
the phase transition happens once for the $S=1/2$ case 
(from one dimerized to the other dimerized phase), and 
twice for the $S=1$ case (from one dimerized to the Haldane phase and 
from the Haldane to the other dimerized phase). 
In the general-$S$ case, there are $2S$ transitions 
from one fully dimerized to the other fully dimerized phase, 
and they are the transitions between the partially dimerized phases.
Around those transition points, the system is represented by the same 
effective field theory as explained in this paper. 

\acknowledgments

The authors appreciate the fruitful discussions with Shunsuke C. Furuya. 
The computation in the present work was partially performed
on computers at the Supercomputer Center, Institute for Solid State Physics, 
University of Tokyo.
This work is also supported in part by
Grants-in-Aid from JSPS, Grant No. 09J08714 (S.T.)
and No. 21540381 (M.O.),
and U.S. NSF Grant No. NSF PHY05-51164
through Kavli Institute for Theoretical Physics,
University of California at Santa Barbara where
part of the present work was completed.


\begin{thebibliography}{}

\bibitem{Giamarchi}
T. Giamarchi, {\it Quantum Physics in One Dimension} 
(Oxford University Press, New York, 2004).

\bibitem{Oshikawa97}
M. Oshikawa and I. Affleck, \PRLp{79}{2883}{1997}.

\bibitem{Affleck99}
I. Affleck and M. Oshikawa, \PRBp{60}{1038}{1999}.

\bibitem{Umegaki11}
I. Umegaki, H. Tanaka, T. Ono, M. Oshikawa and H. Nojiri, 
Physica E {\bf 43}, 
\href{http://www.sciencedirect.com/science/article/pii/S1386947710004327}{741} (2011).

\bibitem{Affleck89}
I. Affleck, D. Gepner, H. J. Schulz and T. Ziman, 
J. Phys. A: Math. Gen. {\bf 22}, 
\href{http://iopscience.iop.org/0305-4470/22/5/015}{511} (1989).

\bibitem{Singh99}
R. R. P. Singh and Z. Weihong, \PRBp{59}{9911}{1999}.

\bibitem{Orignac04}
E. Orignac, Eur. Phys. J. B {\bf 39}, 
\href{http://www.springerlink.com/content/c8yhf08vwrq96eg5/}{335} (2004).

\bibitem{Hase93}
M. Hase, I. Terasaki, and K. Uchinokura, 
\PRLp{70}{3651}{1993}.

\bibitem{Hagiwara98}
M. Hagiwara, Y. Narumi, K. Kindo, M. Kohno, H. Nakano, R. Sato, and M. Takahashi, 
\PRLp{80}{1312}{1998}.

\bibitem{Haldane-NNN}
F. D. M. Haldane, \PRBp{25}{4925}{1982}.

\bibitem{Okamoto92}
K. Okamoto and K. Nomura, Phys. Lett. A {\bf169},
\href{http://www.sciencedirect.com/science/article/pii/0375960192908235}{433} (1992).

\bibitem{Kolezhuk96}
A. Kolezhuk, R. Roth, and U. Schollwock, \PRL{77}{5142}{1996}.

\bibitem{Pati96}
S. Pati, R. Chitra, D. Sen, H. R. Krishnamurthy, and S. Ramasesha,
Europhys. Lett. {\bf 33}, 
\href{http://dx.doi.org/10.1209/epl/i1996-00403-3}{707} (1996).

\bibitem{AffleckHaldane}
I. Affleck and F. D. M. Haldane, \PRBp{36}{5291}{1987}.

\bibitem{Kato94}
Y. Kato and A. Tanaka, \JPSJ{63}{1277}{1994}.

\bibitem{Bouzerar98}
G. Bouzerar, A. P. Kampf, and G. I. Japaridze, \PRBp{58}{3117}{1998}.

\bibitem{Controzzi04}
D. Controzzi and G. Mussardo, \PRL{92}{021601}{2004}.

\bibitem{Vidal07}
G. Vidal, \PRL{98}{070201}{2007}.

\bibitem{Essler98}
F. H. L. Essler and A. M. Tsvelik, \PRBp{57}{10592}{1998}.

\bibitem{Dashen75}
R. F. Dashen, B. Hasslacher, and A. Neveu, 
Phys. Rev. D {\bf 11}, \href{http://prd.aps.org/abstract/PRD/v11/i12/p3424_1}{3424} (1975).

\bibitem{Schulz86}
H. J. Schulz, \PRBp{34}{6372}{1986}.

\bibitem{Zam2_1979}
A. B. Zamolodchikov and Al. B. Zamolodchikov, Ann. Phys. NY {\bf 120}, 
\href{http://www.sciencedirect.com/science/article/pii/0003491679903919}{253} (1979).

\bibitem{Zam2_1992}
A. B. Zamolodchikov and Al. B. Zamolodchikov, Nucl. Phys. B {\bf 379}, 
\href{http://www.sciencedirect.com/science/article/pii/055032139290136Y}{602} (1992).

\bibitem{Haldane83}
F. D. M. Haldane, Phys. Lett. A, {\bf 93}, 
\href{http://www.sciencedirect.com/science/article/pii/037596018390631X}{464} (1983).

\bibitem{Oshikawa92}
M. Oshikawa, J. Phys. Condens. Matter {\bf 4}, 
\href{http://iopscience.iop.org/0953-8984/4/36/019}{7469} (1992).

\bibitem{Pollmann2010}
F. Pollmann, A.~M. Turner, E. Berg, and M. Oshikawa, \PRB{81}{064439}{2010}.

\bibitem{Pollmann2012}
F. Pollmann, E. Berg, A.~M. Turner, and M. Oshikawa, \PRB{85}{075125}{2012}.

\bibitem{Furuya11}
S. C. Furuya, T. Suzuki, S. Takayoshi, Y. Maeda, and M. Oshikawa, 
\PRBR{84}{180410}{2011}.

\bibitem{Kuzmenko09}
I. Kuzmenko and F. H. L. Essler, \PRB{79}{024402}{2009}.

\bibitem{Takayoshi10}
S. Takayoshi and M. Sato, \PRB{82}{214420}{2010}.

\bibitem{Chitra95}
R. Chitra, S. Pati, H. R. Krishnamurthy, D. Sen, and S. Ramasesha,
\PRBp{52}{6581}{1995}.

\bibitem{Kosterlitz74}
J. M. Kosterlitz, J. Phys. C {\bf 7}, 
\href{http://iopscience.iop.org/0022-3719/7/6/005}{1046} (1974).

\end{thebibliography}
\end{document}